\documentclass[12pt,english]{oxarticle}
\pdfoutput=1
\usepackage{graphicx, float, array, xspace, amscd, amsmath, amsthm, amssymb, latexsym, bbm, longtable, mathtools, caption, textcomp, afterpage, xspace, rangecite}
\usepackage[bottom]{footmisc}
\usepackage[all,pdf]{xy}\SelectTips {cm}{}
\usepackage[sort&compress, comma, square, numbers]{natbib}
\usepackage{html}
\usepackage{navigator}
\usepackage[none]{hyphenat}

\let\SS=\S 

\renewcommand{\a}{\alpha}

\renewcommand{\S}{\sum}

\DeclareFontFamily{OT1}{pzc}{}
\DeclareFontShape{OT1}{pzc}{m}{it}{<-> s * [1.200] pzcmi7t}{}
\DeclareMathAlphabet{\mathpzc}{OT1}{pzc}{m}{it}

\newcommand{\IA}{\mathbb{A}}

\newcommand{\IP}{\mathbb{P}}
\newcommand{\IQ}{\mathbb{Q}}

\newcommand{\IZ}{\mathbb{Z}}

\font\eightrm=cmr8 at 8pt
\font\csc=cmcsc10

\DeclareFontFamily{U}{wncy}{}
\DeclareFontShape{U}{wncy}{m}{n}{<->wncyr10}{}
\DeclareSymbolFont{mcy}{U}{wncy}{m}{n}
\DeclareMathSymbol{\sha}{\mathord}{mcy}{"58}

\newcommand{\eu}[1]{{\mathfrak #1}}

\newcommand{\varstr}[2]{\vrule height #1 depth #2 width0pt}

\newcommand{\place}[3]{\vbox to0pt{\kern-\parskip\kern-7pt
                             \kern-#2truein\hbox{\kern#1truein #3}
                             \vss}\nointerlineskip}
                             
\newcommand{\capt}[3]{\parbox{#1}{\renewcommand{\baselinestretch}{1.0}
                                                           \caption{\label{#2}\small\it #3}}}

\newcommand{\K}{K\"{a}hler\xspace}

\newcommand{\beq}{\begin{equation}}
\newcommand{\eeq}{\end{equation}}
\newcommand{\beqnn}{\begin{equation*}}
\newcommand{\eeqnn}{\end{equation*}}

\newcommand{\fref}[1]{Figure~\ref{#1}}
\newcommand{\tref}[1]{Table~\ref{#1}}

\newcommand{\+}{\hphantom{-}}

\newcommand{\cicy}[2]{\begin{matrix} #1\end{matrix}\!\left[\begin{matrix}#2 \end{matrix}\right]}
\newcommand{\gcicy}[3]{\begin{matrix} #1\end{matrix}\!\left[\begin{array}{#3} #2 \end{array} \right]}

\newcommand{\smallgcicy}[3]{{\begin{matrix} #1\end{matrix}\!\left[\!\!\begin{array}{#3} #2 \end{array}\!\!\right]}}

\newcommand{\quotient}[1]{_{\hskip-2pt\lower1pt\hbox{$/$}\lower2pt\hbox{\hskip-1pt$#1$}}}

\newlength{\myht}
\newlength{\mydp}
\newlength{\mywd}
\newsavebox{\mybox}
\newcommand{\entry}[2]{\settowidth{\mywd}{\footnotesize${}\quotient{#2}$}%
\hspace{\mywd}%
\sbox{\mybox}{\footnotesize$#1\quotient{#2}$}%
\settoheight{\myht}{\usebox{\mybox}}\addtolength{\myht}{6pt}%
\settodepth{\mydp}{\usebox{\mybox}}\addtolength{\mydp}{5pt}%
\vrule height\myht width0pt depth\mydp\usebox{\mybox}}

\newcommand{\simpentry}[1]{%
\sbox{\mybox}{$#1$}%
\settoheight{\myht}{\usebox{\mybox}}\addtolength{\myht}{6pt}%
\settodepth{\mydp}{\usebox{\mybox}}\addtolength{\mydp}{5pt}%
\vrule height\myht width0pt depth\mydp\usebox{\mybox}}
\def\str{\vrule height14pt depth8pt width0pt}

\newcommand{\one}{1}            
\addtocounter{secnumdepth}{1}
\addtocounter{tocdepth}{1}
\renewcommand{\baselinestretch}{1.2}
\numberwithin{equation}{section}
\begin{document}
\pagestyle{empty}
\begin{center}
\null\vskip0.3in
{\Huge Calabi-Yau Threefolds\\[20pt] 
with Small Hodge Numbers}\\[0.29in]
{\csc Philip Candelas\footnote{ candelas@maths.ox.ac.uk}, Andrei Constantin\footnote{ andrei.constantin@physics.uu.se}  \\[0.1cm]
and \\[0.04cm]
Challenger Mishra\footnote{ challenger.mishra@gmail.com}\\[1.2cm]}
{\it $^1$Mathematical Institute\hphantom{$^1$}\\
University of Oxford\\
Radcliffe Observatory Quarter\\ 
Woodstock Road
Oxford OX2 6GG, UK\\[2ex]
$^2$Department of Physics and Astronomy\hphantom{$^2$}\\
Uppsala University \\ 
SE-751 20, Uppsala, Sweden\\[2ex]
$^3$Rudolf Peierls Centre for Theoretical Physics\hphantom{$^3$}\\
University of Oxford\\
Clarendon Laboratory, Parks Rd\\
Oxford OX1 3PU, UK\\
}

\vfill
{\bf Abstract\\[3ex]}
\parbox{6.0in}{\setlength{\baselineskip}{14pt}
We present a list of Calabi-Yau threefolds known to us, and with holonomy groups that are precisely $SU(3)$, rather than a subgroup, with small Hodge numbers, which we understand to be those manifolds with height $(h^{1,1}+h^{2,1})\le 24$. With the completion of a project to compute the Hodge numbers of free quotients of complete intersection Calabi-Yau threefolds, most of which were computed in Refs.~\cite{Candelas:2008wb,Candelas:2010ve,candelas2015hodge} and the remainder in Ref.~\cite{constantin2017hodge}, many new points have been added to the tip of the Hodge plot, updating the reviews by Davies and Candelas in Refs.~\cite{Candelas:2008wb,Davies:2011fr}. In view of this and other recent constructions of Calabi-Yau threefolds with small height, we have produced an updated list. }
\end{center}
\newpage
\setcounter{page}{1}
\pagestyle{plain}

\section{Preamble}
\vspace{-10pt}
The tip of the Hodge plot has seen large population growth in recent years. Refs.~\cite{Candelas:2008wb} and~\cite{Davies:2011fr} provided lists of Calabi-Yau threefolds with small Hodge numbers that were comprehensive at the time. The recent papers \cite{candelas2015hodge,constantin2017hodge} computing Hodge numbers of smooth CICY quotients by freely acting symmetries have increased this list considerably. These CY manifolds are in some sense, among the simplest CY manifolds. It is an interesting question to what extent one can base phenomenologically viable string vacua on these spaces. In Ref.~\cite{Braun:2011hd}, an example of a Calabi-Yau manifold with Hodge numbers (1, 1), so the smallest possible for a manifold with a mirror, was obtained.

We list here all the Calabi-Yau threefolds with small Hodge numbers that are known to us (Tables 1-2). For this purpose, small Hodge numbers are all pairs $(h^{1,1},h^{2,1})$ such that $h^{1,1}+h^{2,1}\le 24$, although for most cases we do not list the mirror manifolds. This bound of 24 on the height is arbitrarily chosen, though it coincides with the choice in Ref.~\cite{Candelas:2008wb} and also corresponds to a height low enough so as to separate these points from the Kreuzer-Skarke list. A list of Calabi-Yau manifolds with larger height, albeit somewhat dated, can be found in the online resource~\cite{Jurke}. In this paper, we make no attempt to include Calabi-Yau constructions whose holonomy group is a proper subgroup of $SU(3)$ such as the ones that appear in \cite{andriot2016new}. Neither do we attempt to list orbifolds or manifolds that are derived from tori. 

We also produce an updated version of the tip of the Hodge plot in Figure 1. The Hodge plot becomes denser as we go up in height and merges into the Kreuzer-Skarke list. This list is denser especially owing to the fact that many points have high multiplicities. \fref{TipHodgePlot} indicates many new Hodge numbers found in Refs.~\cite{Candelas:2008wb, Candelas:2010ve, candelas2015hodge,constantin2017hodge}.

There are many methods that were used to construct these Calabi-Yau threefolds. We do not review the methods used to construct this list, since many of these have been discussed in \cite{Candelas:2008wb,Davies:2011fr}. An exception is made for the potentially rich construction introduced in \cite{Anderson:2015iia}. In that work, a generalization of Complete Intersection Calabi-Yau manifolds was presented, wherein some new manifolds, termed gCICYs, were found. Two types of gCICYs that were presented are of codimension 2 and 3 respectively:

\begin{equation*}
X_{(1,1)}= \gcicy{\IP^{n_1}\\ \IP^{n_2}\\ \vdots\\ \IP^{n_m}\\}{a^1&b^1\\a^2&b^2\\\vdots&\vdots\\a^m&b^m}{>{\hskip-2pt}c|c <{\hskip-2pt}},
~~~ X_{(2,1)}= \gcicy{\IP^{n_1}\\ \IP^{n_2}\\ \vdots\\ \IP^{n_m}\\}{a^1_1&a^1_2&b^1\\a^2_1&a^2_2&b^2\\\vdots&\vdots&\vdots\\a_1^m&a_2^m&b^m}{>{\hskip-2pt}cc|c <{\hskip-2pt}}
\end{equation*}

where $a^{i}, a^{i}_1, a^{i}_2 \in \IZ_{\ge0}$ and $b^{i}$ can assume negative integer values. The $``a"$-columns correspond to homogeneous polynomials, i.e.\  globally defined sections of non-negative degree line bundles. On the other hand, the mixed-degree line bundle associated with the $``b"$-column has no global sections on the ambient multi-projective space. However, in special cases, it can have global sections on the sub-manifold defined by the vanishing of the $``a"$-polynomials, thus defining a hypersurface. The polynomial conditions are applied sequentially from left to right. The condition that these matrices define a Calabi-Yau manifold is the same as that for ordinary CICYs i.e., the sum of the entries in the $k^{th}$ row equals $n_k+1$.
\begin{figure}[H]
\begin{center}
\includegraphics[width=6.4in]{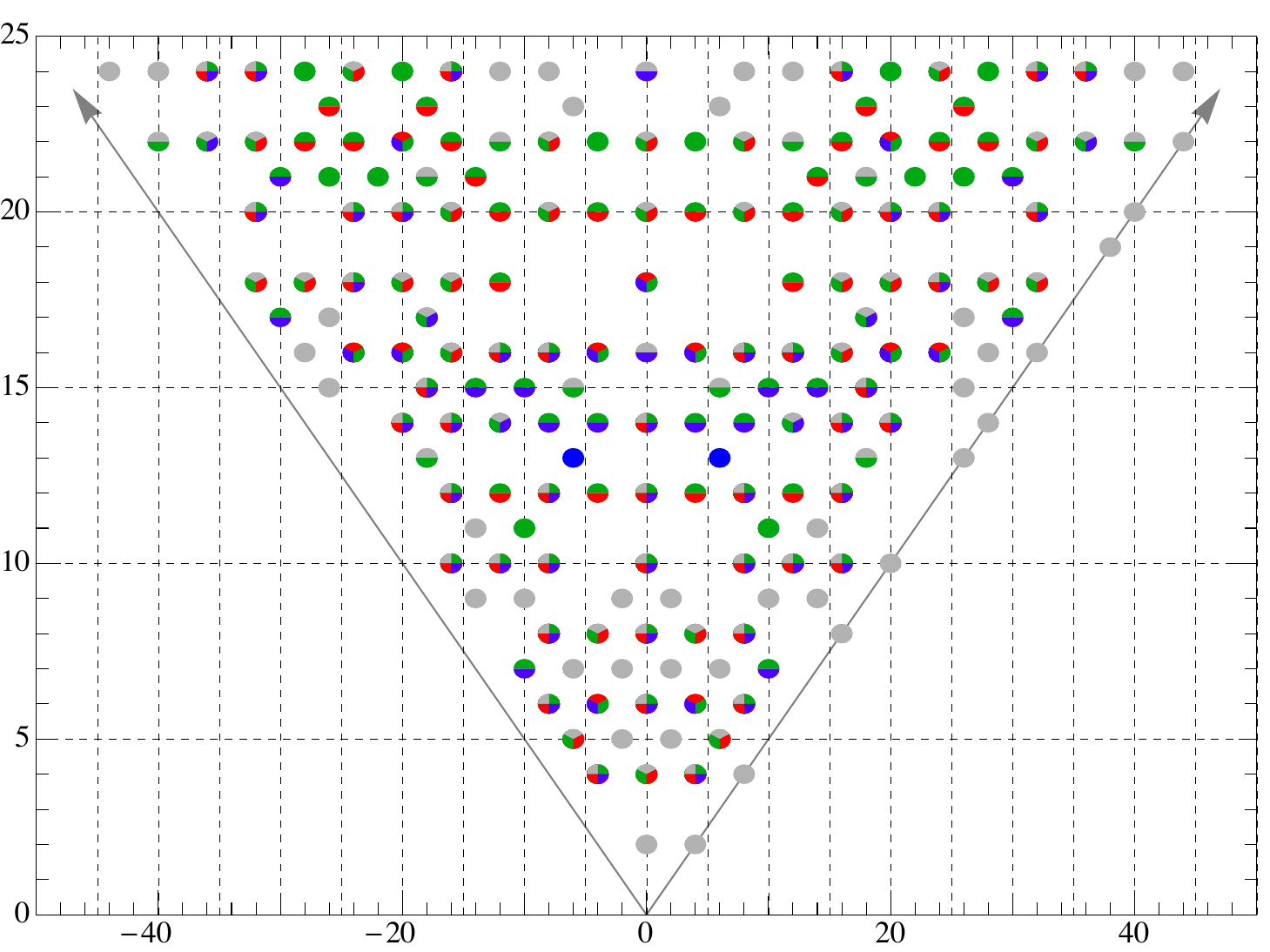}
\vskip18pt
\capt{6.3in}{TipHodgePlot}{The tip of the Hodge number plot for all the Calabi-Yau three-folds that we know, assuming the existence of mirrors except for the rigid manifolds with $h^{2,1}=0$. The grey points are the manifolds of the Kreuzer-Skarke list, CICYs, generalized CICYs, toric CICYs, resolutions of toric conifolds, Gross-Popescu manifolds, the manifold of V.~Braun with Hodge numbers (1,1), manifolds obtained through hyperconifold transitions and other manifolds studied in Refs.~\cite{Yau:1986gu, Green:1987cr, Rodland:1998pm, Kreuzer:2000xy, Gross:2001as, Kreuzer:2003xx, Klemm:2004km, Tonoli:2004ps, Kreuzer:2007ni, Candelas:2007ac, Hua:2007fq, Kapustka:2007pc, Bouchard:2007mf, Batyrev:2008rp, Braun:2009qy, Kapustka:2010pr, Garbagnati:2010um, Stapledon:2010mo, Davies:2011fr, Davies:2011is, Braun:2011hd, freitag2011siegel, filippini2011rigid, Borisov:2012xx, bini2012, Anderson:2015iia, hori2013linear}, as well as the mirrors of the foregoing. The blue points correspond to the CICY quotients (or their desingularisations) studied in \cite{Candelas:2008wb} or \cite{Candelas:2010ve} and their mirrors. The red points correspond to CICY quotients studied in Ref.~\cite{candelas2015hodge} together with their mirrors. The green points correspond to all CICY quotients which were subsequently studied together in Ref.~\cite{constantin2017hodge} and their mirrors. Monochrome points indicate quotients whose Hodge numbers fall onto sites otherwise unoccupied, while the multicoloured points correspond to multiply occupied sites.}
\end{center}
\vskip-10pt
\end{figure}
\newpage
\section{Tables}
Below we give the table of Calabi-Yau manifolds corresponding to \fref{TipHodgePlot}. \tref{BigTable} lists Calabi-Yau threefolds in decreasing order of height $h^{1,1}+h^{2,1}$ and $h^{1,1}$. In our notation, $X^{p,q}$ denotes a manifold with $h^{1,1}=p$ and $h^{2,1}=q$, while $\text{dP}_n$ denotes a del Pezzo surface of degree~$n$. In particular, $X^{20,20}$ refers to the manifold associated with the 24-cell discussed in Ref.~\cite{Braun:2011hd}. $X^{19,19}$ refers to the split bicubic discussed e.g. in \SS 2.2.2 of \cite{Davies:2011fr}. The variety $X^\sharp$ denotes a singular member of a generically smooth family of manifolds and $\widehat{X}$ denotes a resolution of a singular variety $X^\sharp$.  The notation $\IP^7[2~2~2~2]^\sharp$ and $\IP^7[2~2~2~2]^{\sharp\sharp}$, corresponding to an entry with Hodge numbers (1, 17), denote two different singularizations. We denote by $\chi$ and $y$, the Euler characteristic and height $h^{1,1}+h^{2,1}$, of the manifold in question. 

Mirrors of several manifolds with $\chi{\;<\;}0$ have been constructed (exceptions include the manifold by Tonoli \cite{Tonoli:2004ps} and the more recent ones by Kapustka \cite{Kapustka:2010pr}) and we do not list them in our table. The mirrors of toric hypersurfaces and CICYs are known to exist, and may be identified, via the Batyrav and Batyrev-Borisov constructions~\cite{batyrev1993dual, batyrev1996calabi}. For quotients of these toric hypersurfaces and CICYs, the situation is less clear. The mirrors are generally assumed to exist though, as yet, there is no general geometric construction of these mirrors. This assumption of the existence of mirrors applies also to Calabi-Yaus constructed by other means. In \fref{TipHodgePlot}, we assume the existence of the mirrors of the manifolds that we know. 

In each row of \tref{BigTable}, we separate Calabi-Yau constructions studied in the same reference by \textit{commas}, and those studied in distinct references by \textit{semi-colons}. For example the first row refers to three constructions that were studied in Refs.~\cite{Candelas:2008wb}, \cite{Braun:2011hd} and \cite{Anderson:2015iia} respectively, whereas the second row refers to three examples all of which were studied in Ref.~\cite{Anderson:2015iia}.  We group together manifolds with identical Hodge numbers in a single row, unless there are too many to fit, in which case we distribute these over multiple~rows.

At the end, we devote a short table (\tref{TQres}) to (resolved) quotients of resolutions of singular complete intersections of four quadrics in a $\IP^7$ with homogeneous coordinates $(X_0, \ldots,X_3, Y_0,\ldots, Y_3)$. In particular we write $\eu{X}$, for the following complete intersection~\cite{freitag2011siegel}:
\begin{align*}
Y_0^2&=X_0^2+X_1^2+X_2^2+X_3^2\\
Y_1^2&=X_0^2-X_1^2+X_2^2-X_3^2\\
Y_2^2&=X_0^2+X_1^2-X_2^2-X_3^2\\
Y_3^2&=X_0^2-X_1^2-X_2^2+X_3^2~.
\end{align*}
$\eu{X}$ has 96 isolated singularities and admits a resolution which is a Calabi-Yau threefold. The Hodge numbers of the quotients of $\eu{X}$ by groups isomorphic to $\IZ_2^{m}$, for different values of $m$, are listed in the \tref{TQres} along with other resolutions.
\begin{center}
\def\str{\vrule height15pt width0pt depth9pt}
\begin{longtable}{|>{\hskip-4pt}c<{\hskip-4pt}| >{\hskip-5pt}c<{\hskip-5pt} |>{\hskip-3pt}c<{\hskip-3pt} | >{\hskip-6pt}c<{\hskip-6pt} |}
\caption{\label{BigTable} \textit{Table of Calabi-Yau manifolds with small Hodge numbers.}}\\[20pt]
\hline
\str $(\chi,\, y)$ & $(h^{11},h^{21})$ & Manifold & Ref. \\
\hline\hline
\endfirsthead
\multicolumn{4}{c}{\tablename\ \thetable{} \emph{-- Continued from previous page}} \\[1ex]
\hline 
\str $(\chi,\, y)$ & $(h^{11},h^{21})$ & Manifold & Ref. \\
\hline\hline
\endhead
\hline\hline \multicolumn{4}{|r|}{{\str Continued on next page}} \\ \hline
\endfoot
\hline\hline
\multicolumn{4}{|c|}{\str}\\
\hline
\endlastfoot
(0, 24)&(12, 12)& \varstr{45pt}{40pt} \footnotesize \simpentry{\widehat{X^{19,19}\quotient{\IZ_2}}};~~~~~$X^{20,20}/{\IZ_2}$;~~~~$\smallgcicy{\IP^1\\ \IP^1\\ \IP^1\\ \IP^1\\ \IP^1}{1&\+1\\1&\+1\\0&\+2\\0&\+2\\4&-2}{c|>{\hskip-2pt}c}$ & \cite{Candelas:2008wb}; \cite{Braun:2011hd}; \cite{Anderson:2015iia}\\ 
\hline
(-8, 24)&(10, 14)& 
\varstr{45pt}{40pt}\footnotesize
$\smallgcicy{\IP^{2}\\ \IP^{1}\\ \IP^{1} \\ \IP^{1} \\ \IP^{1}\\}{1&0&\+2\\0&1&\+1\\ 0&1&\+1\\1&3&-2\\1&0&\+1}{c>{\hskip-4pt}c|>{\hskip-2pt}c}$,~~~~
$\smallgcicy{\IP^{2}\\ \IP^{1}\\ \IP^{1} \\ \IP^{1} \\ \IP^{1}\\}{1&0&\+2\\0&3&-1\\ 0&1&\+1\\1&1&\+0\\1&0&\+1}{c>{\hskip-4pt}c|>{\hskip-2pt}c}$,~~~~
$\smallgcicy{\IP^{2}\\ \IP^{1}\\ \IP^{1} \\ \IP^{1} \\ \IP^{1}\\}{1&0&\+2\\0&0&\+2\\ 0&1&\+1\\1&1&\+0\\1&3&-2}{c>{\hskip-4pt}c|>{\hskip-2pt}c}$
& \cite{Anderson:2015iia}\\
\hline	
(-12, 24)&(9, 15)& \varstr{35pt}{30pt} \footnotesize$\smallgcicy{\IP^{3}\\ \IP^{1}\\ \IP^{1} \\ \IP^{1} \\}{1&0&\+3\\0&1&\+1\\ 1&1&\+0\\1&3&-2}{c>{\hskip-4pt}c|>{\hskip-2pt}c}$ & \cite{Anderson:2015iia}\\
\end{longtable}
\end{center}
\newpage
\begin{center}
\def\str{\vrule height15pt width0pt depth9pt}
\begin{longtable}{| c | c | c | c |}
\caption{\label{TQres} Hodge numbers of various resolutions of the manifold $\IP^7[2~2~2~2]$.}\\
\hline 
\str $(\chi,\, y)$ & $(h^{11},h^{21})$ & Manifold & Reference \\ 
\hline\hline
\endfirsthead
\multicolumn{4}{c}{\tablename\ \thetable{} \emph{-- Continued from previous page}} \\[1ex]
\hline 
\str $(\chi,\, y)$ & $(h^{11},h^{21})$ & Manifold & Reference \\
\hline\hline 
\endhead
\hline\hline
\multicolumn{4}{|r|}{\str\emph{Continued on the following page}}\\
\hline
\endfoot
\hline\hline
\multicolumn{4}{|c|}{\str}\\
\hline
\endlastfoot
(40, 24)&(22, 2)& \varstr{20pt}{9pt} $\widehat{\widehat{\eu{X}}/ G}$, $G\in\{\IZ_2^{2}, \IZ_2^{4}\}$&  \cite{freitag2011siegel}\\
                  \hline
(32, 24)&(20, 4)& \varstr{20pt}{9pt} $\widehat{\widehat{\eu{X}}/ G}$, $G\in\{\IZ_2, \IZ_2^{3}, \IZ_2^{5}\}$&  \cite{freitag2011siegel}\\
                  \hline
(16, 24)&(16, 8)& \varstr{20pt}{9pt} $\widehat{\widehat{\eu{X}}/ G}$, $G\in\{\IZ_2^{2}, \IZ_2^{4}\}$&  \cite{freitag2011siegel}\\
                  \hline
(44, 22)&(22, 0)& \varstr{20pt}{9pt} $\widehat{\widehat{\eu{X}}/ G}$, $G\in\{\IZ_2, \IZ_2^{3}, \IZ_2^{5}\}$&  \cite{freitag2011siegel}\\
                  \hline
(40, 20)&(20, 0)& \varstr{20pt}{9pt} $\widehat{\widehat{\eu{X}}/ G}$, $G\in\{\IZ_2^{2}, \IZ_2^{4}\}$&  \cite{freitag2011siegel}\\
                  \hline
(32, 20)&(18, 2)& \varstr{20pt}{9pt} $\widehat{\widehat{\eu{X}}/ G}$, $G\in\{\IZ_2, \IZ_2^{3}\}$&  \cite{freitag2011siegel}\\
                  \hline
(16, 20)&(14, 6)& \varstr{20pt}{9pt} $\widehat{\widehat{\eu{X}}/\IZ_2^{2}}$&  \cite{freitag2011siegel}\\
                  \hline
(8, 20)&(12, 8)& \varstr{20pt}{9pt} $\widehat{\widehat{\eu{X}}/\IZ_2^{3}}$&  \cite{freitag2011siegel}\\
\end{longtable}
\vfill
\end{center}
\section*{Acknowledgments}
The work of PC is supported by EPSRC grant BKRWDM00. AC~would like to thank the Mathematical Institute at Oxford University for hospitality during part of the preparation of this paper. CM would like to thank the Rhodes Trust and the Yusuf and Farida Hamied Foundation for partial support during this work. The authors would like to thank Eberhard Freitag and Riccardo Salvati Manni for providing detailed correspondence between the Hodge pairs and groups that appear in \tref{TQres}, and Andreas Braun for helpful discussions. 

\renewcommand{\baselinestretch}{0.9}\normalsize
\bibliographystyle{utcaps}
\bibliography{bibfile}
\end{document}